\documentclass{svjour3}                     
\smartqed  \usepackage{graphicx}
\usepackage{fix-cm}
\usepackage{amsmath}

\journalname{General Relativity and Gravitation}

\begin{document}

\title{Regular models  with quadratic equation of state}

\author{S. D. Maharaj \and P. Mafa Takisa}

\institute{
S. D. Maharaj \and P. Mafa Takisa      \at
Astrophysics and Cosmology Research Unit,
 School of Mathematical Sciences, 
 University of KwaZulu-Natal, Private Bag X54001,Durban 4000, South Africa
\\
\email{maharaj@ukzn.ac.za} \\ \\
}
\date{Received: date / Accepted: date}

\maketitle

\begin{abstract}
We provide new exact solutions to the Einstein-Maxwell system of equations which are physically 
reasonable. The spacetime is static and spherically symmetric with a charged 
matter distribution. We utilise an equation of state which is quadratic 
relating the radial pressure to the energy density. Earlier models, with linear and 
quadratic equations of state, are shown to be contained in our general class of 
solutions. The new solutions to the Einstein-Maxwell are found in terms of 
elementary functions. A physical analysis of the matter and electromagnetic 
variables indicates that the model is well behaved and regular. In particular 
there is no singularity in the proper charge density at the stellar centre unlike 
earlier anisotropic models in the presence of the electromagnetic field.


\end{abstract}

\section{\label{b} Introduction }
The study of charged relativistic objects in general relativity is achieved by solving the Einstein-Maxwell system 
of equations and imposing conditions for physical acceptability. This is not easy to achieve because of the 
nonlinearity of the field equations. The exact solutions found have many applications in relativistic 
astrophysics. The models generated have been used in the description of neutron stars and black hole 
formation by Ray et al.~\cite{1} and de Felice et al.~\cite{2}. Particular models have also helped in the 
establishment of the absolute stability limit for charged spheres by Giuliani et al.~\cite{3} and Bohmer 
and Harko \cite{4}. Several models of charged relativistic matter have been used to study strange stars by 
Mak and Harko \cite{5}, Bombaci \cite{6}, Komathiraj and Maharaj \cite{7} and Thirukkanesh and Maharaj \cite
{8}. Charged models have been also used in the description of strange quark matter by 
Discus et al.~\cite{9}, hybrid protoneutron stars by Nicotra et al.~\cite{10}, and bare quark stars 
by Usov et al.~\cite{11}. A geometric approach  is to assume the existence of a group of conformal motions on spacetime; 
exact solutions have been found by Mak and Harko \cite{12} for strange quark matter and Esculpi and Aloma 
\cite{13} for anisotropic relativistic charged matter by assuming the existence of a conformal killing vector in 
static spherically symmetric spacetimes. 

Models with an equation of state are desirable in the description of realistic astrophysical matter. 
However most explicit solutions of the Einstein-Maxwell system that have been found do not satisfy this 
property. There have been some attempts made recently to find exact analytic solutions of the Einstein-
Maxwell system with a linear equation of state. These include the treatments of Ivanov \cite{14}, Sharma and 
Maharaj \cite{15}, and Thirukkanesh and Maharaj \cite{8}. Particular solutions with a quadratic equation of 
state, relating the radial pressure to the energy, where found by Feroze and Siddiqui \cite{16}. This is an 
important advance since the complexity of the model is greatly increased because of the nonlinearity of the 
radial pressure in terms of the energy density. However the investigations mentioned above all suffer from the 
undesirable property of possessing a singularity in the property charge density at the centre of sphere. An 
essential requirement for a well behaved electromagnetic field is regularity of the proper charge density 
throughout the matter distribution, particularly at the stellar centre. The importance of this feature has been 
highlighted in the analysis of Varela et al.~\cite{17} whose treatment offers a general approach of dealing 
with anisotropic charged matter with linear or nonlinear equations of state. It is desirable to eliminate the 
singularity in the charge density for a detailed and complete analysis of physical properties of charged 
compact objects.

Our results may be helpful in the study of compact stars and gravitational collapse relating to neutron stars and black holes. 
In this regard we refer to particular papers some of which have static spherical geometry and others are dynamical.
 Novikov \cite{17-1} showed in the case of spherical geometry that collapse of electrically charged matter may 
 replaced by expansion and infinite densities are avoided. A general treatment of collapsing charged matter
  was completed by Bekenstein \cite{17-2} who showed that nonzero pressure plays a significant role. 
  The analysis of Raychaudhuri \cite{17-3} for charged dust distributions showed that conditions for 
  collapse and oscillation depend on the ratio of matter density to charge density. If this ratio is large, 
  corresponding to weakly charged dust spheres, then shell crossings cannot be avoided in gravitational 
  collapse as proved by Ori \cite{17-4}. Krasinski and Bolejko \cite{17-5} showed that there exist initial
   conditions for a charged dust sphere with finite radius so that a full cycle of pulsation can be 
   completed by the outer layer with no internal singularity. A full and comprehensive analysis of charged, 
   dissipative collapse is provided by Di Prisco  \textit{et al} \cite{17-6} for the free-streaming and diffusion 
   approximations. A related and detailed analysis in the gravitational collapse of a charged medium 
   was performed by Kouretsis and Tsagas \cite{17-7} where the role of Raychaudhuri equation
   is highlighted. Exact solutions with an equation of state, such the quadratic case considered 
   in this paper, are helpful in such studies.

The objective of this paper is to find new exact solutions of the Einstein-Maxwell field equations with a charged 
anisotropic matter distribution and a quadratic equation of state. We indicate that particular models found in 
the past with an equation of state are part of our general analytical framework. Previous solutions with a linear 
or quadratic equation of state are regained in our treatment. We ensure that the charge density is regular at 
the centre of the compact body and the physical criteria are satisfied. In Sect.~\ref{Basic}, we give the 
Einstein-Maxwell field equations for a static spherically symmetric line element as an equivalent system of 
differential equations utilizing a transformation due to Durgapal and Bannerji \cite{18}. In Sect.~\ref{c1}, we 
present new exact solutions to the Einstein-Maxwell system with a quadratic equation of state. The solution is 
regular at the centre of the compact object. This analysis extends the treatment of Thirukkanesh and Maharaj 
\cite{8}, and Feroze and Siddiqui \cite{16}. Known solutions with an equation of state are presented 
in Sect.~\ref{d}, as particular cases of our new results. In Sect.~\ref{P}, a physical analysis of the new solutions is 
performed; the matter variables and the electromagnetic quantities are plotted. 
We summarise the results obtained in this paper.

\section{Field equations\label{Basic}}

In standard coordinates the line element for a static spherically symmetric spacetime, 
modelling the interior of the relativistic object, has the form
\begin{equation}
\label{f1} ds^{2} = -e^{2\nu(r)} dt^{2} + e^{2\lambda(r)} dr^{2} 
+ r^{2}(d\theta^{2} + \sin^{2}{\theta} d\phi^{2}).
\end{equation}
We take the energy momentum tensor to be of the form
\begin{equation}
\label{f100}
T_{ij}= \mbox{diag} (-\rho-\frac{1}{2}E^{2}, ~ p_{r} - \frac{1}{2}E^{2}, ~ p_{t}
 + \frac{1}{2}E^{2},  ~ p_{t} + \frac{1}{2}E^{2}),
\end{equation}
where the quantity $p_{t}$ is the tangential pressure, $p_{r}$ is the radial
 pressure, $\rho$ is the density,  and $E$ is the electric field intensity.
Then the Einstein-Maxwell equations can be written in the form
\begin{eqnarray}
\label{f2}
\dfrac{1}{r^{2}}\big[r(1-e^{-2\lambda})\big]^{\prime} &=& \rho+\dfrac{1}{2}E^{2},\\
\label{f3}
-\dfrac{1}{r^{2}}(1-e^{-2\lambda})+\dfrac{2\nu^{\prime}}{r}e^{-2\lambda} &=& p_{r}-\dfrac{1}{2}E^{2},\\
\label{f4}
e^{-2\lambda}\bigg(\nu^{\prime\prime}+\nu^{\prime2}+\dfrac{\nu^{\prime}}{r}
-\nu^{\prime}\lambda^{\prime}-\dfrac{\lambda^{\prime}}{r}\bigg) &=& p_{t}+\dfrac{1}{2}E^{2},\\
\label{f5}
\sigma &=& \dfrac{1}{r^{2}}e^{-\lambda}(r^{2}E)^{\prime},
\end{eqnarray}
where primes represent differentiation with respect to \textit{r}. The quantity 
$\sigma$ represents the proper charge density. We are utilising units where the coupling 
constant $\frac{8\pi G}{c^{2}}=1$ and the speed of light $c=1$. The mass within a radius \textit{r} of the sphere is 
\begin{equation}
\label{f101}
M(r)=\frac{1}{2}\int^{r}_{0}\omega^{2}\rho(\omega)d\omega.
\end{equation}
We now introduce a new independent variable \textit{x} and define new functions \textit{y} and \textit{Z} so that
\begin{equation}
\label{f6} x = Cr^2,~~ Z(x)  = e^{-2\lambda(r)}, ~~
A^{2}y^{2}(x) = e^{2\nu(r)},
\end{equation}
where \textit{A} and \textit{C} are constants. We assume an equation of state
 of the general form $p_{r} = p_{r}(\rho)$ for the matter distribution.
  We take the quadratic form
\begin{equation}
p_{r} = \gamma\rho^{2}+\alpha\rho-\beta  \label{f7},
\end{equation}
relating the radial pressure $p_{r}$ to the energy density $\rho$. In the
 above $\alpha$, $\beta$, and $\gamma$ are constants. Then the Einstein-Maxwell 
 equations governing the gravitational behaviour of a charged anisotropic sphere, 
 with a quadratic equation of state, are represented by
\begin{eqnarray}
\label{f8}
\dfrac{\rho}{C} &=& \frac{1-Z}{x}-2\dot{Z}+\dfrac{E^{2}}{2C},\\
\label{f9}
p_{r} &=& \gamma\rho^{2}+\alpha\rho-\beta,\\
\label{f10}
p_{t} &=& p_{r}+\Delta,\\
\label{f11}
\frac{\Delta}{C} &=& 4xZ \frac{\ddot{y}}{y}+2[x\dot{Z}+2Z]\frac{\dot{y}}{y}-
\alpha \left[\frac{(1-Z)}{x}-2\dot{Z}-\frac{E^{2}}{2C} \right]\nonumber\\
& & -C\gamma\left[\frac{(1-Z)}{x}-2\dot{Z}-\frac{E^{2}}{2C} \right]^{2}
+\dot{Z}-\frac{E^{2}}{2C}+\frac{\beta}{C},\\
\label{f12}
\dfrac{\dot{y}}{y} &=& \frac{(1-Z)(1+\alpha)}{4Z}-\frac{(1+\alpha)E^{2}}{8CZ}-
\frac{\alpha\dot{Z}}{4Z}-\frac{\beta}{4CZ}\nonumber\\
& & +\frac{C\gamma}{4Z}\left[\frac{(1-Z)}{x}-2\dot{Z}-\frac{E^{2}}{2C} \right]^{2},\\
\label{f13}
\dfrac{\sigma^{2}}{C} &=& \dfrac{4Z}{x}\left(x\dot{E}+E\right)^{2},
\end{eqnarray}
where dots denote differentiation with respect to the variable $x$. 
Equations (\ref{f8})-(\ref{f13}) are similar to the field equations of Thirukkanesh 
and Maharaj \cite{8}; however in this case the equation of state is quadratic.
The quantity $\Delta = p_t -p_r$ is called the measure of anisotropy and vanishes 
for isotropic pressures. The nonlinear system as given in (\ref{f8})-(\ref{f13}) consists 
of six independent equations in six variables involving the matter and electromagnetic 
quantities $\rho,  p_r,  p_t, \Delta,  E, \sigma$ and the two gravitational potentials \textit{y} 
and \textit{Z}. The nonlinearity of the Einstein-Maxwell system (\ref{f8})-(\ref{f13}) 
has been increased, when compared with many earlier treatments, become 
of the appearance of the quadratic term in (\ref{f9}); when $\gamma = 0$ then 
there is a linear equation of state. In addition equation (\ref{f12}) now contains 
terms with $E^{4}$ increasing the complexity of system since $\gamma \neq 0$ in general.

\section{\label{c1} New solutions} 

To integrate the Einstein-Maxwell system we make the particular choices
\begin{eqnarray}
Z &=&\frac{1+bx}{1+ax}, \label{S1}\\
\frac{E^{2}}{C}&=&\dfrac{k(3+ax)+sa^{2}x^{2}}{(1+ax)^{2}}. \label{S2}
\end{eqnarray}
The gravitational potential \textit{Z} is well behaved and finite at the origin. 
The electric field intensity \textit{E} is continuous, regular at the origin and 
approaches a constant value for increasing values of \textit{x}. The constants
 \textit{a}, \textit{b}, \textit{k} and \textit{s} are real. The general analytic 
 functional forms for \textit{Z} and \textit{E} regain particular cases studied 
 in the past with an equation of state.

On substituting (\ref{S1}) and (\ref{S2}) into (\ref{f12}) we obtain the first order equation
\begin{eqnarray}
\frac{\dot{y}}{y} &=& 
 \dfrac{(1+\alpha)(a-b)}{4[1+(a-b)x]}+\dfrac{{\alpha}(a-b)}{2(1+ax)[1+bx]}-
 \frac{\beta(1+ax)}{4C[1+bx]}\nonumber\\
 & & -\frac{(1+\alpha)[k(3+ax)+sa^{2}x^{2}]}{8(1+ax)[1+bx]}+
 \frac{C\gamma[(3+ax)(2a-ab-k)-sa^{2}x^{2}]^{2}}{16(1+ax)(1+bx)}, \label{S4}
\end{eqnarray}
for the metric function \textit{y}. In spite of complexity of equation 
(\ref{S4}) it can be solved in general.
On integrating $(\ref{S4})$ we get
\begin{equation}
y=D(1+ax)^{m}[1+bx]^{n}\exp\left[F(x)\right],\label{S5}
\end{equation}
where \textit{D} is the constant of integration. The function \textit{F(x)} 
is given explicitly by
\begin{eqnarray}
F(x) &=& \gamma [2(a-b)-k]^{2}\left[\frac{2(2b-a)(1+ax)+(b-a)}{8(b-a)^{2}(1+ax)^{2}}\right]\nonumber\\
&& -Cs\gamma\left[ \frac{(a-b)^{2}(ax+2)-a(2a+s)(1+ax)}{4(a-b)(1+ax)^{2}}\right]\nonumber\\
&& -Cs\gamma\left[\frac{(a-b)(4k+s)+(2k(b-3a)+3bs)(1+ax)}{32(a-b)^{2}(1+ax)^{2}} \right]\nonumber\\
& & +\frac{ax}{16bC}[C^{2}s^{2}\gamma -2Cs(1+\alpha)-4\beta].\label{SF}
\end{eqnarray}
The constants \textit{m} and \textit{n} have the form
\begin{eqnarray}
m &=& -\frac{(1+\alpha)(s+2k)}{8(b-a)}+\frac{\alpha}{2}+\gamma  [2(a-b)-k]^{2}
\left[\frac{b^{2}}{(b-a)^{3}}+\frac{b}{(b-a)^{2}}+ \frac{1}{4}\right]\nonumber\\
& & +\frac{Cs\gamma}{8(a-b)^{3}}\left[(a-b)[2s(a-b)+a+b]+3ab(k-2b)-b^{2}k\right.\nonumber\\
&&\left. +2b^{3}(2a-1) \right]\nonumber\\
n &=& \frac{(1+\alpha)}{8b}[2(a-b)-k]+\frac{(1+\alpha)k-2\alpha(a-b)}{4(b-a)}
+ \frac{\beta (a-b)}{4Cb^{2}}\nonumber\\
&& +\gamma  [2(a-b)-k]^{2}\left[\frac{b^{2}}{(b-a)^{3}}+\frac{b}{(b-a)^{2}}+ 
\frac{1}{4}\right]+ \frac{Csa^{2}(1+\alpha)}{8b^{2}(b-a)}\nonumber\\
&& +\frac{Cs\gamma}{16b^{2}(b-a)^{3}}\left[a^{4}(s+4b)+(k+2b)(6a^{2}b^{2}-2a^{3}b)\right].\nonumber
\end{eqnarray}
Then we can generate an exact model for the Einstein-Maxwell system (\ref{f8})-(\ref{f13}) in the form 
\begin{eqnarray}
\label{S6}
e^{2\lambda} &=& \frac{1+ax}{1+bx},\\
\label{S7}
e^{2\nu} &=& A^{2}D^{2}\left(1+ax \right)^{2m}[1+bx]^{2n}\exp[2F(x)],\\
\label{S8}
\dfrac{\rho}{C} &=& \dfrac{(2a-2b-k)(3+ax)-sa^{2}x^{2}}{2(1+ax)^{2}},\\
\label{S9}
p_{r} &=& \gamma \rho^{2}+\alpha \rho-\beta ,\\
\label{S10}
p_{t} &=& p_{r}+\Delta,\\
\label{S11}
\frac{\Delta}{C} &=&  \frac{4x(1+bx)}{1+ax}\left[\frac{m(m-1)a^{2}}{(1+ax)^{2}}
+\frac{2mnab}{(1+ax)(1+bx)} + \frac{2ma\dot{F}(x)}{1+ax}+\right.\nonumber\\
& & \left. \frac{b^{2}n(n-1)}{(1+bx)^{2}} + \frac{2nb\dot{F}(x)}{1+bx}+ \ddot{F}(x)+ \dot{F}(x)^{2}\right]
+\left[-\frac{2(a-b)x}{(1+ax)^{2}}+\dfrac{4(1+bx)}{(1+ax)}\right]\nonumber\\ 
& &\left[\frac{am}{1+ax}+ \frac{bn}{1+bx} + \dot{F}(x) \right]-C\gamma\left[\frac{C(2(a-b)
-k)(3+ax)-Csa^{2}x^{2}}{2(1+ax)^{2}}\right]^{2}\nonumber\\
& & -\frac{1}{2(1+ax)^{2}}[2C(a-b)+k(3+ax)+sa^{2}x^{2}-\frac{2\beta}{C}(1+ax)^{2}]\nonumber\\
& & -\alpha\left[\frac{(2(a-b)-k)(3+ax)-sa^{2}x^{2}}{2(1+ax)^{2}}\right],\\
\label{S12}
\frac{E^{2}}{C}&=& \frac{k(3+ax)+sa^{2}x^{2}}{(1+ax)^{2}},\\
\label{S13}
\frac{\sigma^{2}}{C} &=& \frac{C[1+bx]\left(\sqrt{k}(a^{2}x^{2}+3ax+6)
+2\sqrt{s}ax\sqrt{3+ax}(2+ax)\right)^{2}}{x(3+ax)(1+ax)^{5}},
\end{eqnarray}
where \textit{F(x)} is given by (\ref{SF}). We observe that the exact solution (\ref{S6})-(\ref{S13}) of the 
Einstein-Maxwell system has been written solely in terms of elementary functions. For this solution the mass function is given by
\begin{eqnarray}
\label{S14}
 M(x) &=&  \frac{1}{8C^{3/2}}\left[\frac{[(12a(a-b)-6ak)x+s(15+10ax-2a^{2}x^{2})]x^{1/2}}{3a(1+ax)}\right.\nonumber\\
& & \left.-\frac{5s\arctan(\sqrt{ax})}{a^{3/2}}\right].
\end{eqnarray}
The gravitational potentials, matter variables and electromagnetic variables are well behaved and regular in the stellar interior.
However in general there is a singularity in the charge density at the centre which is evident in (\ref{S13}). 
This singularity is avoidable when $k=0$, so that we have
\begin{equation}
\label{S15}
\frac{\sigma^{2}}{C} = \frac{4Csa^{2}x[1+bx](2+ax)^{2}}{(1+ax)^{5}}.
\end{equation}
 At the centre of the star $x=0$ and the charge density vanishes.

\section{\label{d} Known solutions}
We have found a general class of exact solutions to the Einstein-Maxwell system with a quadratic equation of 
state. It is interesting to observe that for particular parameter values we
can regain uncharged anisotropic and isotropic models $(k=0$, $s=0)$ from
our general solution (\ref{S6})-(\ref{S13}).  We regain the
following particular cases of physical interest.

\subsection{\label{d5} Feroze and Siddiqui model}
This is a special case of our general solution with the quadratic equation of state $p_{r}$
 = $\gamma \rho^{2}+\alpha \rho - \beta$. If we set $s=0$, $C=1$ and $A^{2}D^{2}=B$, then we regain the line element 
\begin{eqnarray}
ds^{2}&=& B(1+ar^{2})^{m}(1+br^{2})^{n}\exp[2F(r)]dt^{2}+\frac{1+ar^{2}}{1+br^{2}}dr^{2}\nonumber\\
&&  +r^{2}(d\theta^{2}+\sin^{2}\theta d\phi^{2})\label{S19},
\end{eqnarray}
where
\begin{eqnarray}
F(x) &=& -\frac{\beta ar^{2}}{4b}+\gamma [2(a-b)-k]^{2}\left[\frac{2(2b-a)(1+ar^{2})+(b-a)}{8(b-a)^{2}(1+ar^{2})^{2}}\right],\nonumber\\
m &=& \frac{\alpha}{2}-\frac{(1+\alpha)k}{4(b-a)}+\gamma  [2(a-b)-k]^{2}\left[\frac{b^{2}}{(b-a)^{3}}+\frac{b}{(b-a)^{2}}+ \frac{1}{4}\right],\nonumber\\
n &=& \frac{(1+\alpha)}{8b}[2(a-b)-k]+\frac{(1+\alpha)k-2\alpha(a-b)}{4(b-a)}+ \frac{\beta (a-b)}{4b^{2}}\nonumber\\
&& +\gamma  [2(a-b)-k]^{2}\left[\frac{b^{2}}{(b-a)^{3}}+\frac{b}{(b-a)^{2}}+ \frac{1}{4}\right].\nonumber
\end{eqnarray}
The line element (\ref{S19}) was found by Feroze and Siddiqui \cite{16} which
 was the first model with quadratic equation of state. Some minor misprints in \cite{16} 
 have been corrected in our result. This solution may be used to model a compact body.

\subsection{\label{d3}Thirukkanesh and Maharaj model}
If we set $\gamma=0$ then we have the linear equation of state $p_{r}$ = $\alpha \rho - \beta$. 
Also setting $C=1$, $s=0$ and $b=a-\tilde{b}$, we get the line element 
\begin{eqnarray}
ds^{2}&=& A^{2}D^{2}(1+ar^{2})^{2m}[1+(a-\tilde{b})r^{2}]^{2n}\exp\left[ \frac{-a\beta r^{2}}{2(a-\tilde{b})}\right]dt^{2}\nonumber\\
&& +\frac{1+ar^{2}}{1+(a-\tilde{b})r^{2}}dr^{2}+r^{2}(d\theta^{2}+\sin^{2}\theta d\phi^{2})\label{S18},
\end{eqnarray}
where
\begin{eqnarray}
m &=& \frac{2 \alpha \tilde{b} -(1+\alpha)k}{4\tilde{b}},\nonumber\\
n&=& \frac{1}{8\tilde{b} (a-\tilde{b})^2} \left[2a^2 (k(1+\alpha)-2\alpha \tilde{b}) -
a \tilde{b} (5k(1+\alpha) -2\tilde{b} (1+5\alpha))\right.\nonumber\\
&& \left.+\tilde{b}^2 (3k(1+\alpha) - 2\tilde{b}
(1+3\alpha)+2\beta)\right].\nonumber
\end{eqnarray}
The metric (\ref{S18}) was found by Thirukkanesh and Maharaj \cite{8}. This solution may be
 used to model realistic charged compact spheres and strange stars with quark matter 
 in the presence of electromagnetic field.

\subsection{\label{d2}Sharma and Maharaj model}
If we set $\gamma=0$, $\beta = \alpha\tilde{\rho}$, then we regain the linear equation
 of state $p_{r}=\alpha(\rho -\tilde{\rho})$ where $\tilde{\rho}$ is the density at the surface.
 By setting $k=0$ $s=0$, $b=a-\tilde{b}$, $C=1$ and $A^{2}D^{2}=B$ we find the following form of the line element
\begin{eqnarray}
ds^{2}&=&-B(1+ar^{2})^{2m}[1+(a-\tilde{b})r^{2}]^{2n}\exp\left( \frac{-a\beta r^{2}}{2(a-\tilde{b})}\right)dt^{2}\nonumber\\
& & +\frac{1+ar^{2}}{1+(a-\tilde{b})r^{2}}dr^{2}+r^{2}(d\theta^{2}+\sin^{2}\theta d\phi^{2})\label{S17},
\end{eqnarray}
 where
 \begin{eqnarray}
m &=& \frac{\alpha}{2},\nonumber\\
n&=&\frac{5a\tilde{b}\alpha-2a^{2}\alpha-3\tilde{b}^{2}\alpha+a\tilde{b}-\tilde{b^{2}}+\tilde{b}\beta}{4(a-\tilde{b})^{2}}. \nonumber
\end{eqnarray}
The line element (\ref{S17}) represents an uncharged anisotropic sphere and was found by
 Sharma and Maharaj \cite{15}. It may be used to describe strange stars with a linear equation of state with quark matter.

\subsection{\label{d1}Lobo model}
If we set $\gamma=0$, $\beta=0$, then we obtain the linear equation of state $p_r=\alpha \rho$.
On setting $k=0$, $s=0$, $b=a-\tilde{b}$, $a=2\tilde{b}$, $C=1$ and $A^{2}D^{2}=B$ we regain the line element 
\begin{eqnarray}
\label{S16}
ds^{2}&=&-(1+2\tilde{b}r^{2})^{2m}(1+\tilde{b}r^{2})^{2n}dt^{2}+\left(\frac{1+2\tilde{b}r^{2}}{1+\tilde{b}r^{2}}\right)dr^{2}\nonumber\\
& & +r^{2}(d\theta^{2}+\sin^{2}\theta d\phi^{2}).
\end{eqnarray}
 where
\begin{eqnarray}
m &=&\frac{\alpha}{2},\nonumber\\
n&=&\frac{1-\alpha}{4}. \nonumber
\end{eqnarray}
The metric (\ref{S16}) was first found by Lobo \cite{19} which represents uncharged anisotropic 
matter. This solution serves as a stellar interior with $\alpha < -\frac{1}{3}$ and may be
 matched to the Schwarzchild exterior for dark energy stars.
\subsection{\label{d6}Isotropic models}
We observe that $\Delta \neq 0$ in general and the model remains anisotropic. However, we can 
show for particular parameter values that $\Delta=0$ in the general solution (\ref{S6})-(\ref{S13}). 
If we set $b=(a-1)$, $k=0$, $s=0$, $a=0$,  then we obtain
\begin{eqnarray}
m &=& \frac{\alpha}{2}\nonumber\\
n &=& \frac{1}{4C}[\beta - (1+3\alpha)C]\nonumber\\
\Delta &=& \frac{x}{4C(1-x)}[\beta-3(1+\alpha)C][\beta-(1+3\alpha)C].\label{S20}
\end{eqnarray}
 Two different cases arise from (\ref{S20}) by setting $\Delta=0$. Firstly, we observe that 
 when $\beta=0$ and $\alpha=-1$ then $\Delta=0$. The equation of state becomes $p_{r}(=p_{t})=-\rho$. With the line element
\begin{eqnarray}
ds^{2}=-\left(1+\frac{r^{2}}{R^{2}}\right)dt^{2}+\left(1+\frac{r^{2}}{R^{2}}\right)^{-1}dr^{2}+r^{2}(d\theta^{2}+\sin^{2}\theta d\phi^{2})\label{S21}
\end{eqnarray}
where we have set $A = D = 1$ and $ C=\frac{1}{R^{2}}$. 
We mention that the metric (\ref{S21}) is the isotropic uncharged de Sitter model. Secondly, 
we observe that when $\beta=0$ and $\alpha=-\frac{1}{3}$ then $\Delta=0$. The equation
 of state becomes $p_{r}(=p_{t})=-\frac{1}{3}\rho$ with the line element
\begin{eqnarray}
ds^{2}=-A^{2}dt^{2}+\left(1+\frac{r^{2}}{R^{2}}\right)^{-1}dr^{2}+r^{2}(d\theta^{2}+\sin^{2}\theta d\phi^{2})\label{S22}
\end{eqnarray}
where $D=1$ and $C=\frac{1}{R^{2}}$.
The metric (\ref{S22}) is the isotropic uncharged Einstein model.

\begin{figure}
\vskip .2cm \centering
\includegraphics[angle = 0,scale = 0.75]{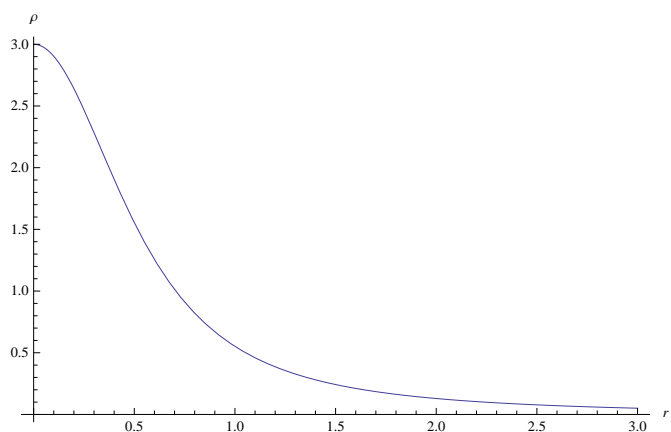}
\caption{Energy density}
\label{fig1}
\end{figure}

\begin{figure}
\vskip .2cm \centering
\includegraphics[angle = 0,scale = 0.75]{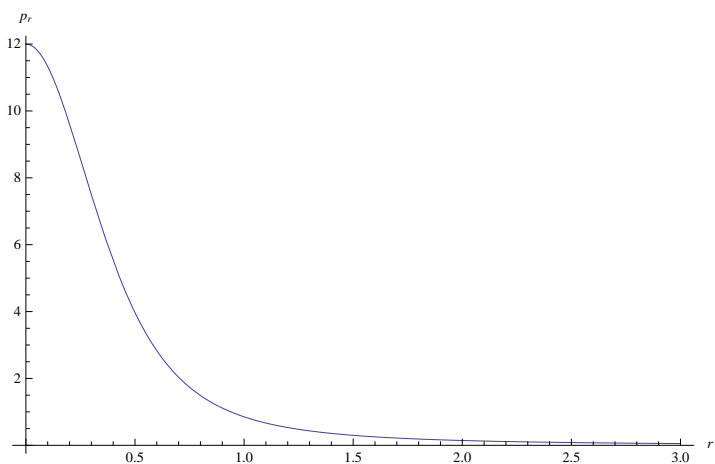}
\caption{Radial pressure}
\label{fig2}
\end{figure}

\begin{figure}
\vskip .2cm \centering
\includegraphics[angle = 0,scale = 0.75]{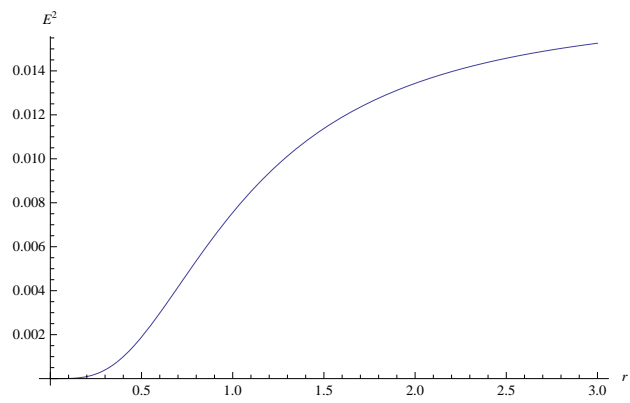}
\caption{Electric field intensity}
\label{fig3}
\end{figure}

\begin{figure}
\vskip .2cm \centering
\includegraphics[angle = 0,scale = 0.95]{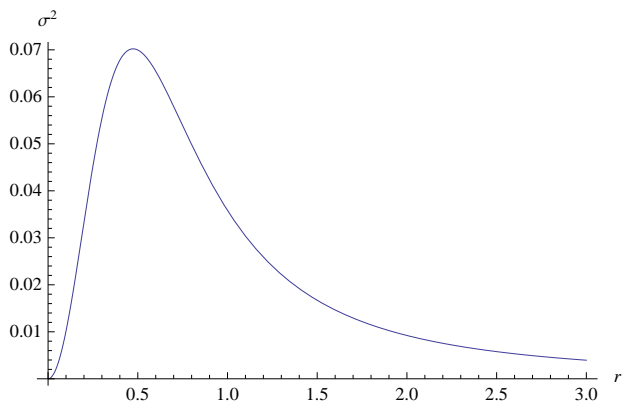}
\caption{Charge density}
\label{fig4}
\end{figure}

\begin{figure}
\vskip .2cm \centering
\includegraphics[angle = 0,scale = 0.75]{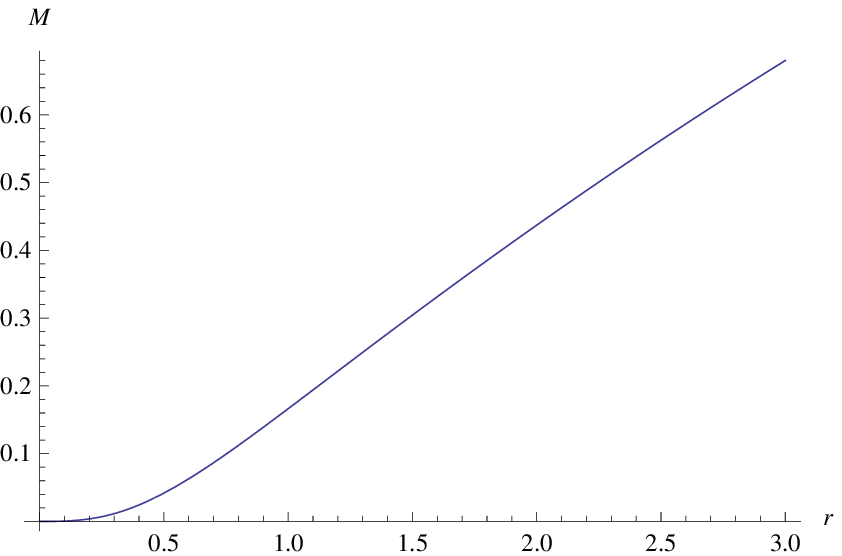}
\caption{Mass}
\label{fig5}
\end{figure}

\begin{figure}
\vskip .2cm \centering
\includegraphics[angle = 0,scale = 0.75]{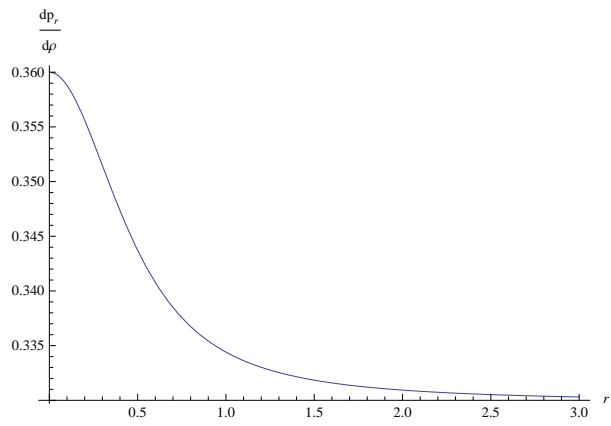}
\caption{Speed of sound}
\label{fig6}
\end{figure}

\begin{figure}
\vskip .2cm \centering
\includegraphics[angle = 0,scale = 0.75]{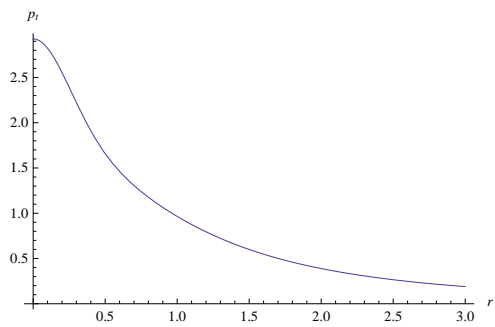}
\caption{Tangential pressure}
\label{fig7}
\end{figure}

\begin{figure}
\vskip .2cm \centering
\includegraphics[angle = 0,scale = 0.75]{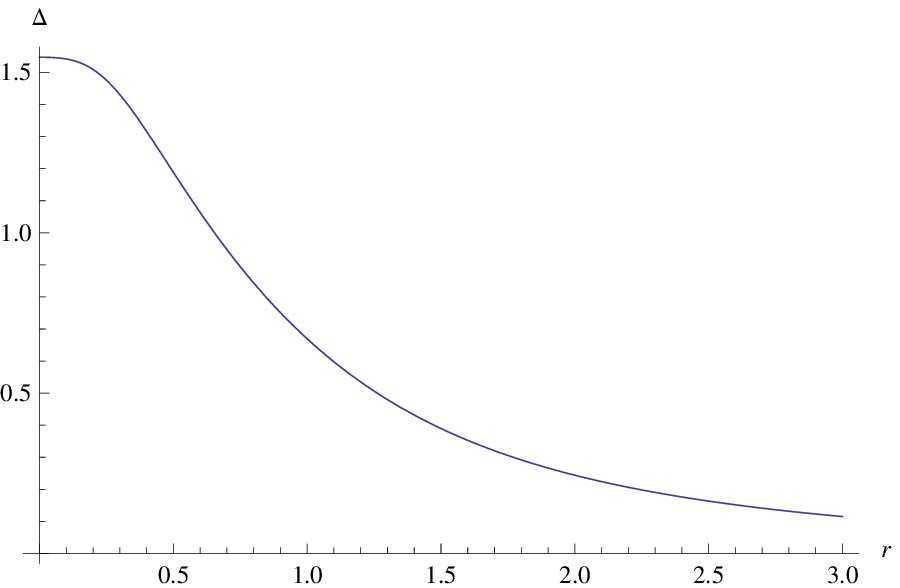}
\caption{Measure of anisotropy}
\label{fig8}
\end{figure}

\section{\label{P} Physical Analysis}
We show that the exact solutions of the Einstein-Maxwell system found in section 3 are well behaved by
 generating graphical plots of matter and electromagnetic variables. We used the software 
 package Mathematica \cite{20} and we make the particular choices 
 $C=1$, $a=2.5$, $b=2$, $\gamma = 0.01$, $\alpha=0.33$, $s=0.017$ and $k=\beta=0$. 
 We generated the plots for the energy density (Fig. 1), radial pressure (Fig. 2), electric field 
 intensity (Fig. 3), charge density (Fig. 4), mass (Fig. 5), speed of sound (Fig. 6), tangential pressure (Fig. 7) 
 and the measure of anisotropy (Fig. 8). The energy density $\rho$ is a finite and monotonically
  decreasing function. The radial pressure $p_{r}$ is similarly well behaved and continuous. The 
  electric field intensity $E$ is initially small and approaches a maximum value as the boundary 
  is approached. The proper charge density $\sigma$ is nonsingular at the origin, increases and then 
  decreases after reaching a maximum value. The mass function is a strictly increasing function which
   is continuous and finite. The speed of sound is less than the speed of light and causality is 
   maintained throughout the stellar interior. The radial pressure is decreasing and does reach 
   a finite value of the radial coordinate. The tangential pressure is also a decreasing function. 
   The measure of anisotropy is a decreasing function as the boundary is approached and
    remains finite in the interior. Thus all the matter variables, electromagnetic variables and
     the gravitational potentials are nonsingular and regular in the region containing the stellar
      centre. In particular the proper charge density $\sigma$ is finite at the centre unlike earlier treatments.

It is desirable to study comprehensively the stability of our new models; this is a objective
 for the future reseach. The solutions generated may be matched to the exterior Reissner-Nordstrom spacetime 
\begin{eqnarray}
ds^{2}&=&-\left(1-\frac{2M}{r}+\frac{Q^{2}}{r^{2}}\right)dt^{2}+\left(1-\frac{2M}{r}+\frac{Q^{2}}{r^{2}}\right)^{-1}dr^{2}\nonumber\\
& & +r^{2}(d\theta^{2}+\sin^{2}\theta d\phi^{2})\label{S104},
\end{eqnarray}
across the boundary $r=\Re$. This generates the following conditions
\begin{eqnarray}
1-\frac{2M}{\Re}+\frac{Q^{2}}{\Re^{2}}=A^{2}y^{2}(C\Re^{2})
\label{S105},
\end{eqnarray}
and
\begin{eqnarray}
\left(1-\frac{2M}{\Re}+\frac{Q^{2}}{\Re^{2}}\right)^{-1}=\frac{1+aC\Re^{2}}{1+bC\Re^{2}}
\label{S106},
\end{eqnarray}
relating the constants \textit{ a, b, A, C,} $\alpha$, $\beta$ and $\gamma$. There 
are  sufficients number of free parameters to ensure the continuity of the metric coefficients across the boundary of the star.
It is possible to study the astrophysical significance of the
 exact solutions to the Einstein-Maxwell equations found in this paper. 
 This is the object of future research. We point out that for suitable parameter 
 values we regain the mass $M = 1.433 M_{\bigodot}$ of Dey $et~ al$ \cite{21}-\cite{23}
  corresponding to a strange star model when there is no electromagnetic field. 
  Therefore the solutions found in this paper may be used to generalise earlier
   results and to model charged relativistic strange and quark stars. 

Our aim in this paper was to find new regular exact solutions to the 
Einstein-Maxwell system for spherically symmetric gravitational field 
with an eq of state. In particular we selected a quadratic equation of state 
relating the energy density to the radial pressure. The new models presented in
 this paper may be used to model relativistic compact objects in astrophysics.

\begin{acknowledgements}
SDM and PMT thank the National Research
Foundation and the University of KwaZulu-Natal for financial
support. SDM acknowledges that this work is based upon research
supported by the South African Research Chair Initiative of the
Department of Science and Technology and the National Research
Foundation. We are grateful to the referees for valuable comments.
\end{acknowledgements}

\end{document}